\documentclass[aps,floatfix]{revtex4}
\usepackage{amsmath}
\usepackage{latexsym}
\usepackage{float}
\usepackage{amssymb}
\usepackage{graphicx}
\usepackage{textcomp}
\usepackage{hyperref}
\usepackage{color}
\newcommand{\bea}{\begin{eqnarray}}
\newcommand{\eea}{\end{eqnarray}}
\newcommand{\vect}[1]{\mathbf{#1}}
\newcommand{\req}{\rho_{\rm eq}}
\newcommand{\mex}{\mu^{\rm ex}}

\begin{document}

\title{Curvature Dependence of Surface Free Energy of Liquid Drops and Bubbles: A Simulation Study}

\author{
        Benjamin J. Block$^{1}$,
          Subir K. Das$^{2,1}$,
          Martin Oettel $^{3,1}$,
          Peter Virnau $^{1}$, and
          Kurt Binder $^{1}$}

\affiliation{$^{1}$ \textit{Institut f\"ur Physik, Johannes Gutenberg-Universit\"at, Staudinger Weg 7, D-55099 Mainz, Germany}\\
                $^{2}$ \textit{Theoretical Sciences Unit, Jawaharlal Nehru Centre for Advanced Scientific Research, Jakkur P.O., Bangalore 560064, India} \\
                {$^{3}$ \textit{Material- und Prozesssimulation,
Universit{\"a}t Bayreuth, N{\"u}rnberger Stra{\ss}e 38, D-95448 Bayreuth }}
}

\date{\today}

\begin{abstract}
We study the excess free energy due to phase coexistence of fluids
by Monte Carlo simulations using successive umbrella sampling in
finite $L\times L \times L$ boxes with periodic boundary
conditions. Both the vapor-liquid phase coexistence of a simple
Lennard-Jones fluid and the coexistence between A-rich and B-rich
phases of a symmetric binary (AB) Lennard-Jones mixture are
studied, varying the density $\rho$ in the simple fluid or the
relative concentration $x_A$ of $A$ in the binary mixture,
respectively. The character of phase coexistence changes from a spherical
droplet (or bubble) of the minority phase (near the coexistence
curve) to a cylindrical droplet (or bubble) and finally (in the
center of the miscibility gap) to a slab-like configuration of two
parallel flat interfaces. Extending the analysis of 
M. Schrader, P. Virnau, and K. Binder [Phys. Rev. E{\bf79},
061104 (2009)], we extract the surface free energy $\gamma (R)$ of
both spherical and cylindrical droplets and bubbles in the
vapor-liquid case, and present evidence that for $R \rightarrow
\infty$ the {leading order (Tolman) correction for droplets has sign opposite to
the case of bubbles, consistent with the Tolman length being independent on the sign of curvature}. For the symmetric binary mixture the
expected non-existence of the Tolman length is confirmed. In all
cases {and for a range of radii} $R$ relevant for nucleation theory,
$\gamma(R)$ deviates strongly from $\gamma (\infty)$ {which can be accounted
for by}
a term of order $\gamma
(\infty)/\gamma(R)-1 \propto R^{-2}$. {Our results for the simple Lennard-Jones fluid are
also compared to results from density functional
theory and we find qualitative agreement in the behavior of $\gamma(R)$ as well as
in the sign and magnitude of the Tolman length.} 
\end{abstract}
\maketitle

\section{Introduction}
Curved interfaces between coexisting vapor and liquid phases (or
between coexisting A-rich and B-rich phases of binary (A,B) liquid
mixtures having a miscibility gap) are ubiquitous in nature.
Nanoscopic spherical droplets or bubbles need to be considered in
the context of nucleation phenomena \cite{1,2,3,4}. In nanoscopic
slit pores or cylindrical pores, wetting or drying phenomena
\cite{5,6,7,8,9} typically can cause a curvature of interfaces
that extend across the pore \cite{10,11,12,13,14}. Note that
phase coexistence in porous media is relevant for widespread
applications \cite{10,15,16,17}, ranging from oil recovery out of
porous rocks to devices in microfluidics.

Understanding the properties of such curved interfaces is a
longstanding and difficult problem. Note that on the atomistic
scale interfaces between coexisting phases are rather diffuse and
{the problem of understanding} their structure is not fully solved. From
the point of view of thermodynamics, the central quantity of
interest is the interfacial tension $\gamma$ and its dependence on
the radius of curvature $R$ of the droplet (or bubble).
Although this problem was already mentioned by
Gibbs \cite{18} and discussed in classical papers by Tolman
\cite{19}, the subject is still controversial. Tolman introduced
\cite{19} a length $\delta (R)$, referred to as ``Tolman's
length'', to describe the curvature dependence of $\gamma(R)$, but
the understanding of
this length is incomplete till now
\cite{20,21,22,23,24,25,26,27,28,29,30,31,32,Kog98,Nap01,33,34,35,36,37,38,39,40}.

When one considers a droplet coexisting with surrounding vapor,
different definitions of the droplet radius are conceivable. One
of them is the ``equimolar radius'' $R$, which assumes that the
volumes of the two phases $V_I$, $V_{II}$ and their particle
numbers $N_I$, $N_{II}$ are additive,
\begin{equation} \label{eq1}
V=V_I+V_{II}, \quad N=N_I+N_{II},
\end{equation}
such that there is neither an excess volume nor an excess particle
number associated with the interface. This definition implies that
the ``interfacial adsorption'' is identically zero and in
equilibrium both phases $I$ (liquid) and $II$ (vapor) have the
same chemical potential $\mu_I=\mu_{II}=\mu$, since they can
exchange particles. For a spherical droplet (or bubble)
we have $V_I=4 \pi R^3/3$, of course, and in the
grand-canonical ensemble with $\mu$ and temperature $T$ as control
variables, the densities of the coexisting phases
$\rho_I(\mu,T)=N_I/V_I$, $\rho_{II} (\mu, T)=N_{II}/V_{II}$ are
those of bulk liquid in equilibrium and surrounding metastable
vapor. Here we disregard the conceptual problem that metastable
states in the framework of statistical mechanics are not
completely well-defined \cite{3,41}.

Another important concept to introduce a division
between the two phases is the ``surface of tension'',
i.e., the surface where the surface tension acts
\cite{18,19,20,21}. Consider, for fixed $R$ as defined above, the
surface tension $\gamma (R_p)$ \cite{18,19,20,21,41}, i.e., the excess
contribution to the thermodynamic potential, as a function of the
radius $R_p$ where we put the dividing surface.
One expects that $\gamma(R_p)$ will exhibit a minimum for a choice
of $R_p$ somewhere in the region of the atomistically diffuse
interface, but in general there is no principle that requires that
$R_p$ and $R$ exactly coincide. Considering the pressure
difference $\Delta p$ between the coexisting phases, one can
derive a generalized Laplace equation \cite{20}
\begin{equation} \label{eq2}
\Delta p= 2 \gamma (R_p)/R_p + \partial \gamma (R_p)/\partial R_p.
\end{equation}
Thus at the ``surface of tension'' where $\partial \gamma
(R_p)/\partial R_p=0$ the pressure difference takes its standard
macroscopic form, $\Delta p= 2 \gamma / R_p$.

Now the proper definition of the Tolman length can be written as
\begin{equation} \label{eq3}
\delta = \lim\limits_{R \rightarrow \infty} \delta (R) =
\lim\limits_{R \rightarrow \infty} (R-R_p).
\end{equation}
Recall that in this treatment $R_p$ and hence $\delta (R)$ in
general can depend on $R$. Tolman \cite{19} also
suggested the approximation that $\delta$ (for a liquid droplet)
is a positive constant. If this is assumed, one can show that the
curvature-dependent surface tension becomes
\begin{equation} \label{eq4}
\gamma(R)=\gamma (\infty)/(1+2 \delta/R).
\end{equation}

On the other hand, there is evidence from density functional calculations
\cite{29,30,Kog98,Nap01} for liquid droplets surrounded by supersaturated
vapor that actually {$\delta(R)$} varies strongly with $R$, even
changing its sign from a positive value at small $R$ to a small
negative value at large $R$. Apart from very recent indications \cite{39,40},
most simulations, some of which are still
inconclusive (see the discussion in \cite{38}),
did not support this result.
{(Note that the positive values for $\delta$ in some previous simulations have
not been obtained by using Eq.~(\ref{eq4}) but by a ``virial route'' through
$1/R$--corrections to the pair correlation function of a {\em planar} interface
\cite{27,38}.)}
There are also
compelling arguments \cite{22,26} that for systems obeying a
strict symmetry between the two coexisting phases, e.g., the
Ising lattice gas model of a fluid that exhibits particle/hole
symmetry, the Tolman length $\delta$ must be identically zero,
since interchanging the identity of the coexisting phases turns a
droplet into a bubble which simply means a change of sign of
the radius of curvature of the interface separating them. This
implies that $\gamma(R)$ for such symmetric systems can only be a
function of $R^2$, invalidating Eq.~(\ref{eq4}) even for
arbitrarily large $R$.

{
There appears to be consensus about the uniqueness of the $1/R$ form of the leading correction to the surface tension 
of bubbles and droplets  (given by the Tolman length) when analyzed in different frameworks. This is not so
for the next--to--leading correction and even for the leading correction to the surface tension
of a cylindrical interface \cite{26, Hen84}. Nevertheless, a phenomenological expansion of the surface tension of an
arbitrarily curved surface in powers of its curvatures  
can be devised using the Helfrich form. For spherical (subscript $s$) and cylindrical interfaces (subscript $c$) the
following expressions are obtained \cite{31}
\bea
 \label{eq:gams}
 \gamma_s (R) & = & \gamma (\infty) - 2 \gamma (\infty) \frac{\delta}{R} + (2k+\bar{k}) \frac{1}{R^2} \;, \\
 \label{eq:gamc}
 \gamma_c (R) & = & \gamma (\infty) - \gamma (\infty) \frac{\delta}{R} + \frac{k}{2} \frac{1}{R^2} \;. 
\eea
Here, $k$ is the bending rigidity constant and $\bar{k}$ is the rigidity constants associated with Gaussian curvature.
Such a form neglects possible nonanalytic terms in $R$, and its usefulness should be judged by comparison to actual results.   
}

In the present work, we make an attempt to study {the
problem of the Tolman length and of higher order corrections}, both by analyzing computer simulations of simple models
for fluids and fluid mixtures, and by density functional theory.
The distinctive features of our work are that we apply a recent
``thermodynamic'' method \cite{39} based on exploiting
Eq.~(\ref{eq1}) for various finite volumes $V=L^3$ with
periodic boundary conditions throughout and calculate
$\gamma(R)$ directly via application of successive umbrella sampling
methods \cite{42} to obtain accurate estimates for the appropriate
thermodynamic potential of our model systems. In subsequent sections 
we use the symbols $\gamma_{AB}$ for surface tension along the $A-B$ interface
in a binary mixture and $\gamma_{vl}$ for vapor-liquid interfacial tension
in a single component system.

The paper is organized in the following sequence. To clarify the 
curvature dependence in symmetrical situations, in Sec. II we 
present results from the study of a symmetric 
binary (A,B) Lennard Jones mixture whose
equilibrium properties have been extensively studied in earlier
works \cite{43,44}.
In Sec. III, we turn
attention to the vapor-to-liquid transition of the Lennard-Jones
fluid, considering both droplets and bubbles on an equal footing.
Previous works, with other methods, have considered droplets almost
exclusively. However, our simulation setup also allows us to
consider cylindrical droplets or bubbles, stabilized by the
periodic boundary conditions. Such cylindrical interfaces are not
only of interest to provide further constraints on the possible
value of $\delta$ as defined in Eq.~(\ref{eq3}), but are also
important when one considers phase coexistence in slit pores. Sec.
IV discusses the problems from the point of view of density
functional theory, while Sec. V summarizes our conclusions.

\section{The Curvature-Dependent Surface Tension in a Symmetric Binary
Lennard-Jones Mixture}

\begin{figure}[htb]
\includegraphics*[width=0.6\textwidth]{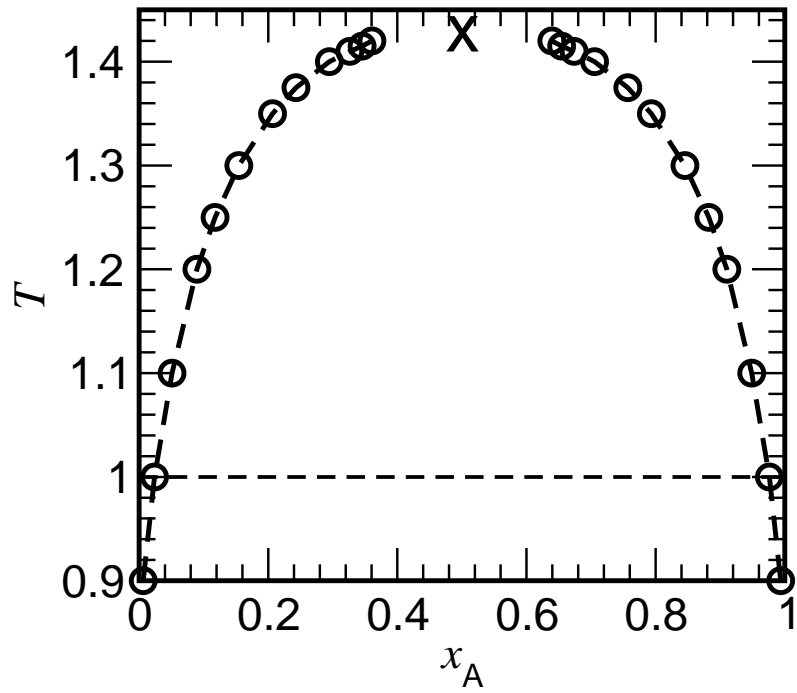}
\caption{Phase diagram of the symmetric binary (A,B) Lennard-Jones
mixture, cf. Eqs.~(\ref{eq5})-(\ref{eq8}), 
in the plane of variables $T$ and
relative concentration of A particles ($x_A=N_A/N$; $N=N_A +
N_B$) for fixed $N=6400$. The cross shows the critical point, as
obtained previously \cite{44}. The horizontal broken line means that
phase coexistence is studied at
$T=1.0$.}\label{fig1}
\end{figure}

We study a binary fluid of $N$ point particles (labeled by index
$i$, at positions $\vec{r}_i$ in the cubical box of finite volume
$V=L^3$ subject to periodic boundary conditions) interacting with
pairwise potentials $u(r_{ij})$ with
$r_{ij}=|\vec{r}_i-\vec{r}_j|$. Starting from a full Lennard-Jones
potential
\begin{equation} \label{eq5}
\phi_{LJ}(r_{ij})=4 \varepsilon_{\alpha \beta} [(\sigma_{\alpha
\beta}/r_{ij})^{12} - (\sigma_{\alpha \beta}/r_{ij})^6], \quad
\alpha, \beta \in A,B,
\end{equation}
we construct a truncated potential as follows \cite{44}:
\begin{equation} \label{eq6}
u(r_{ij})=\phi_{LJ} (r_{ij})-\phi_{LJ} (r_c) - (r_{ij} -r_c)
\frac{d \phi_{LJ}}{d r_{ij}} \mid_{r_{ij}=r_c}, \mbox{for}~ r_{ij} \leq r_c,
\end{equation}
while $u(r_{ij} \geq r_c)=0$. This form ensures that both the
potential and the force are continuous at $r=r_c$ \cite{45}. The
potential parameters are chosen as
\begin{equation} \label{eq7}
\sigma_{AA}=\sigma_{BB}=\sigma_{AB}=\sigma , \quad r_c=2.5 \sigma;
\end{equation}
\begin{equation} \label{eq8}
\varepsilon_{AA}=\varepsilon_{BB}=2 \varepsilon_{AB}=\varepsilon.
\end{equation}
Choosing a reduced density $\rho^*=1$, where
\begin{equation} \label{eq9}
\rho^* =\rho \sigma ^3 =N \sigma^3 /V,
\end{equation}
we work with a dense fluid and at the temperatures of interest 
neither the vapor-liquid transition nor crystallization is a
problem. The unit of temperature $T$ is chosen such that
$\varepsilon/k_B \equiv 1$, also throughout the paper
we set the unit of length $\sigma$ to unity. The critical temperature $T_c$ of phase
separation into an A-rich phase and a B-rich phase occurs at
\cite{44} $T=1.4230 \pm 0.0005$. Using Monte Carlo methods in the
semi-grandcanonical ensemble and applying a finite-size scaling
analysis \cite{46,46p}, the phase diagram in the plane of variables
($T,x_A=N_A/N)$ has been obtained rather accurately in the critical
region \cite{44} and we have extended it here to lower
temperatures as depicted in Fig.\ref{fig1}.

\begin{figure}[htb]
\includegraphics*[width=0.6\textwidth]{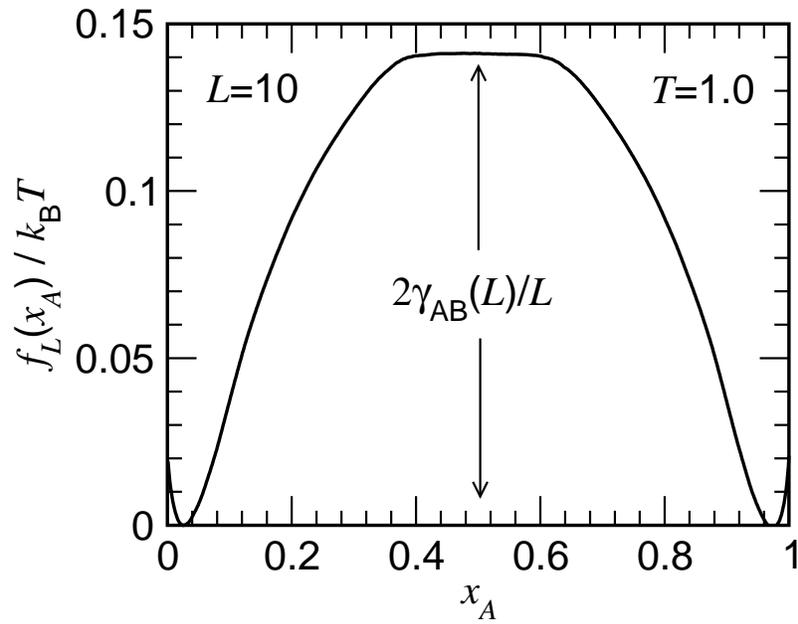}
\caption{Effective free energy $f_L(x_A,T)$ of finite-size boxes
of linear dimension $L$ with $L=10$ plotted vs. $x_A$ at $T=1.0$,
for the model of Fig.~\ref{fig1}. The estimation of the
size-dependent interfacial tension $\gamma_{AB}(L)$ is
indicated.}\label{fig2}
\end{figure}

\begin{figure}[htb]
\includegraphics*[width=0.6\textwidth]{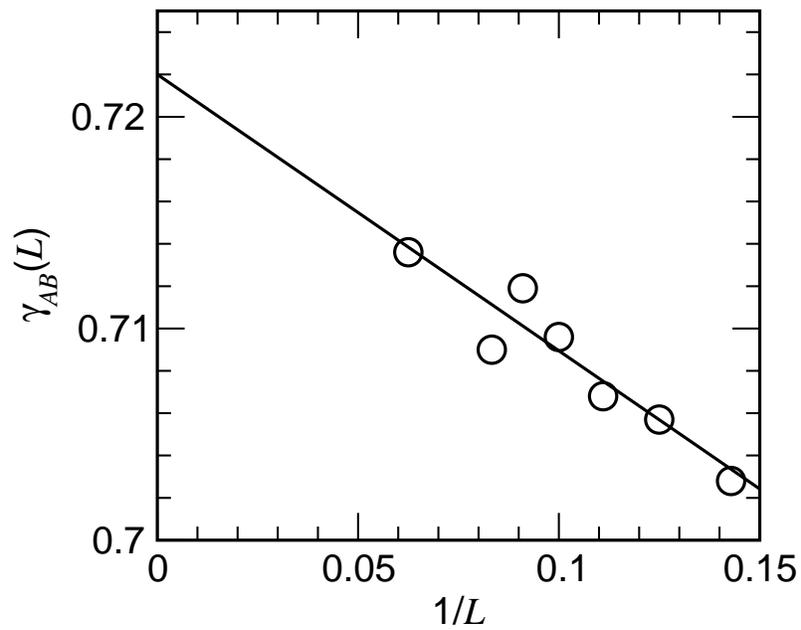}
\caption{Extrapolation of $\gamma_{AB}(L)$ as function of $1/L$,
cf. Eq.~(\ref{eq10}), in order to estimate $\gamma_{AB}(\infty)$.
}\label{fig3}
\end{figure}

The first task then is to obtain reliable estimates for
$\gamma_{AB}=\gamma_{AB}(\infty)$, the interfacial tension between flat,
infinitely extended coexisting A-rich and B-rich phases. Here we
follow the standard method \cite{47,48,49,50,51,52,53,54,55} to
first sample the effective free energy $Vf_L(x_A,T)=-k_B T \ln [P_{\Delta
\mu NVT} (x_A)/P_{\Delta \mu NVT} (x^{\rm coex}_A)]$,
presented in Fig.\ref{fig2}, by successive
umbrella sampling in the semi-grandcanonical ensemble where the
chemical potential difference $\Delta \mu$ between A and B
particles is independently chosen. Note that
the flat region (hump) of $f_L(x_A, T)$ in
Fig.~\ref{fig2} can be interpreted as being 
caused by the excess free energy of two interfaces, of area $L^2$ each, oriented
parallel to two axes
of the cubic box, separating the A-rich domain from the B-rich
domain, and hence
\begin{equation} \label{eq10}
\gamma_{AB} (\infty) =\lim\limits_{L \rightarrow \infty} \gamma
_{AB} (L), \,\, \gamma _{AB} (L)\equiv Lf_L (x_A \approx0.5)/2.
\end{equation}
Note that in finite boxes the capillary wave spectrum
\cite{5,6,7,8,20} of the interfaces is truncated, and there may be
additional corrections due to the translational entropy of the
interfaces, residual interactions between them, etc. Therefore, an
extrapolation of $\gamma_{AB}(L)$ to $L \rightarrow \infty$ is
indeed important to reach a meaningful accuracy. Such exercise is
shown in
Fig.\ref{fig3} where $\gamma_{AB}(\infty)=0.722$ is obtained from a 
linear extrapolation as a function of inverse linear dimension $L$ of systems.

\begin{figure}[htb]
\includegraphics*[width=0.6\textwidth]{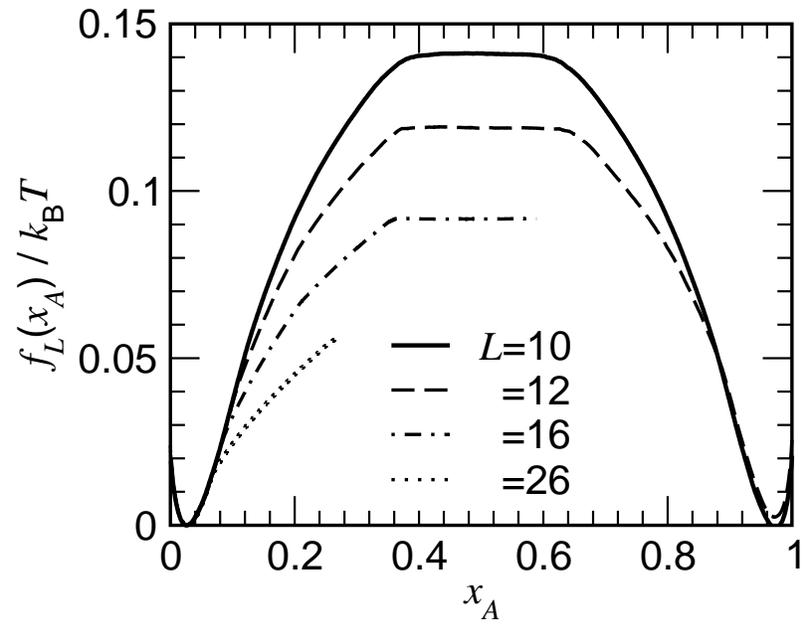}
\caption{Effective free energy $f_L(x_A,T)$ plotted vs. $x_A$ at
$T=1.0$ for $L=10,12,16,26$, as indicated.}\label{fig4}
\end{figure}

\begin{figure}[htb]
\includegraphics*[width=0.5\textwidth]{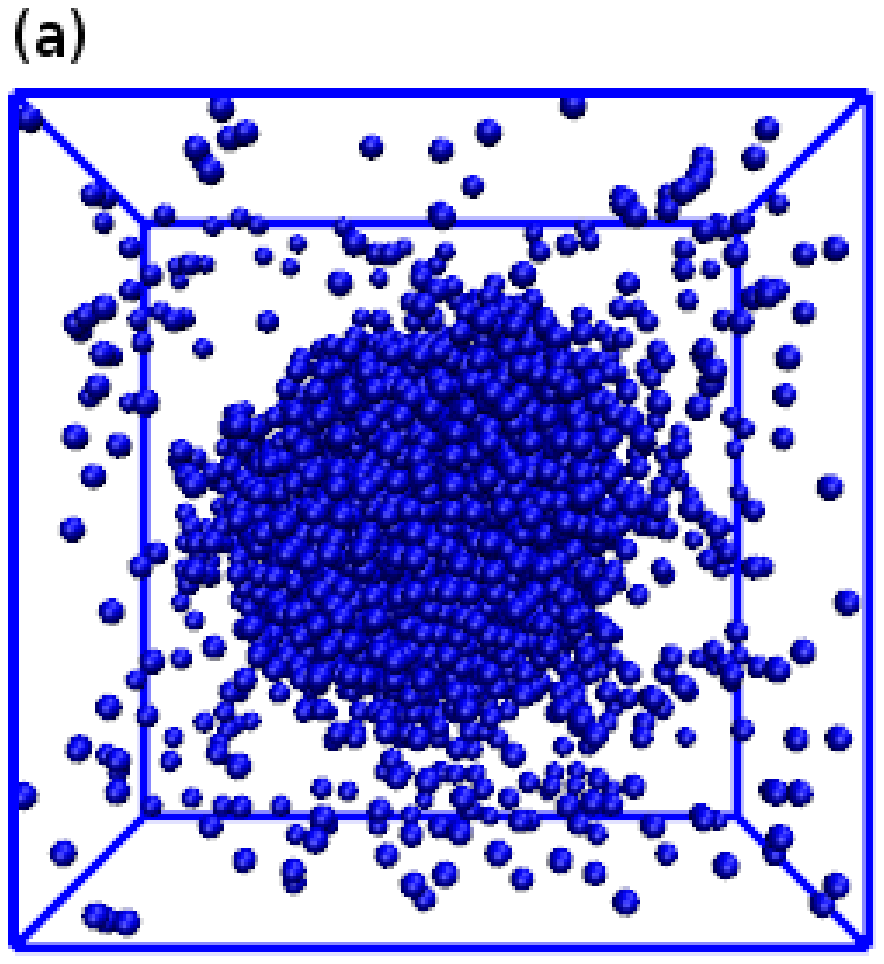}\\
\includegraphics*[width=0.5\textwidth]{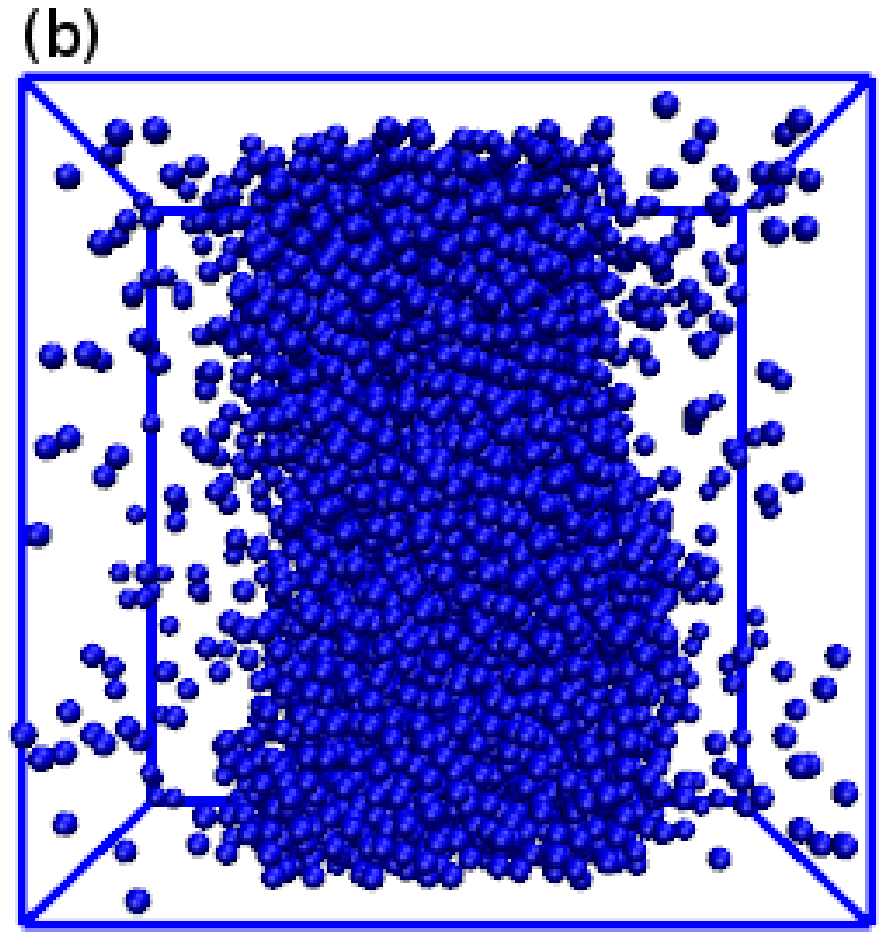}
\caption{a) Snapshot of a spherical droplet configuration formed
by A particles in the background of B-particles (not shown), for
$T=1.0$, $L=24$, and $x_A=0.15$. b) Same as (a), but for a
cylindrical droplet, choosing $x_A=0.27$. 
Note that our method does not at all suppress statistical
fluctuations in the size and shape of these droplets, which therefore
have spherical or cylindrical symmetry on the average only.
}\label{fig5}
\end{figure}

\begin{figure}[htb]
\includegraphics*[width=0.6\textwidth]{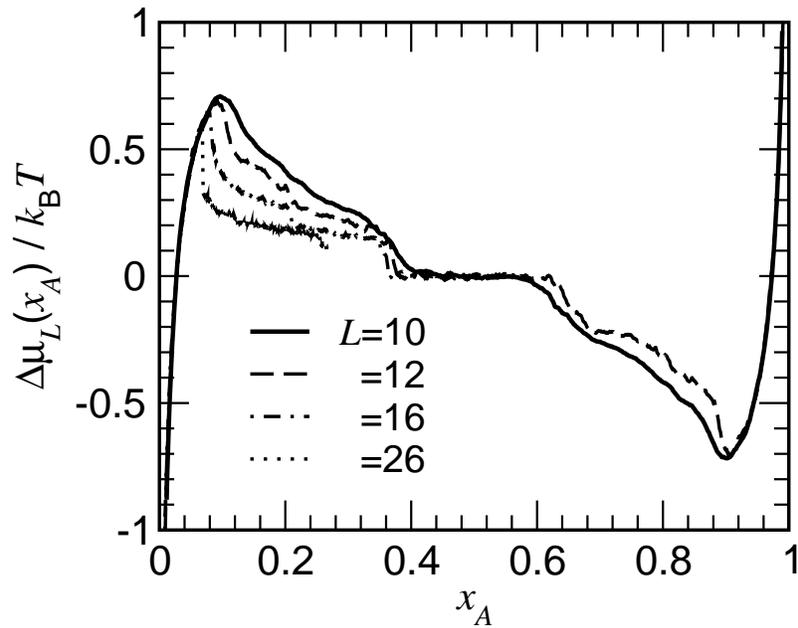}
\caption{Plot of $\Delta \mu(X_A)/k_BT$ vs. $X_A$ at $T=1.0$ for
$L=10,12,16,26$. Data refer to a single run at each size, to
illustrate the typical noise level ($200$ Monte Carlo steps per
particles have been used for each window of the successive
umbrella sampling). For the final analysis, $5$ such runs were
averaged over.} \label{fig6}
\end{figure}

As described in the work of Schrader et al. \cite{39} and Winter
et al. \cite{55,56}, the curvature dependant surface tension $\gamma _{AB} (R)$ 
is extracted from $f_L(x_A,T)$ using other parts of the same curve
in Fig.~\ref{fig2}, namely parts that reflect the
coexistence of an A-rich droplet with B-rich background. Such
states are found in the ascending, size-dependent part of
$f_L(x_A,)$, seen in Fig.~\ref{fig4}, and correspond to spherical droplets
for smaller $x_A$ and cylindrical droplets for larger $x_A$, representative
snapshots of which are shown in
Fig.~\ref{fig5}. In order
to be able to analyze $f_L(x_A,T)$ and distinguish what fraction
of this effective free energy is due to the surface free energy of
the droplet and what is due to the background, a B-rich phase
supersaturated in A-particles analogous to a vapor phase in the gas-liquid context,
we also
need to study the effective chemical potential difference $\Delta
\mu_L(x_A)$: 
\begin{equation} \label{eq11}
\frac{1}{k_BT} \Delta \mu_L (x_A)=[\partial f_L (x_A, T)/\partial
x_A]_T\;, 
\end{equation}
{which we present in Fig.~\ref{fig6}.}
One can clearly identify
the various transitions: the peak in the $\Delta \mu_L(x_A)$ vs.
$x_A$ curve for small $x_A$ is the transition from the homogeneous
state of the box to droplet + ``vapor'' coexistence (the so-called
droplet evaporation/condensation transition
\cite{39,55,56,57,58,59,Tro05}); the next step (near $x_A \approx 0.2)$
signifies the transition of the droplet from spherical to
cylindrical shape; and finally, near $x_A \approx 0.35$, the
transition from the cylinder to the slab configuration occurs for which
we have $\Delta \mu_L(x_A)=0$, of course. In order to
extract information on droplet surface free energies only those
parts of the $f_L(x_A,T)$ vs. $x_A$ and $\Delta \mu_L (x_A, T)$
vs. $x_A$ curves can be used that are not at all affected by
fluctuations associated with these transitions
\cite{39,55,56,57,58}.

\begin{figure}[htb]
\includegraphics*[width=0.6\textwidth]{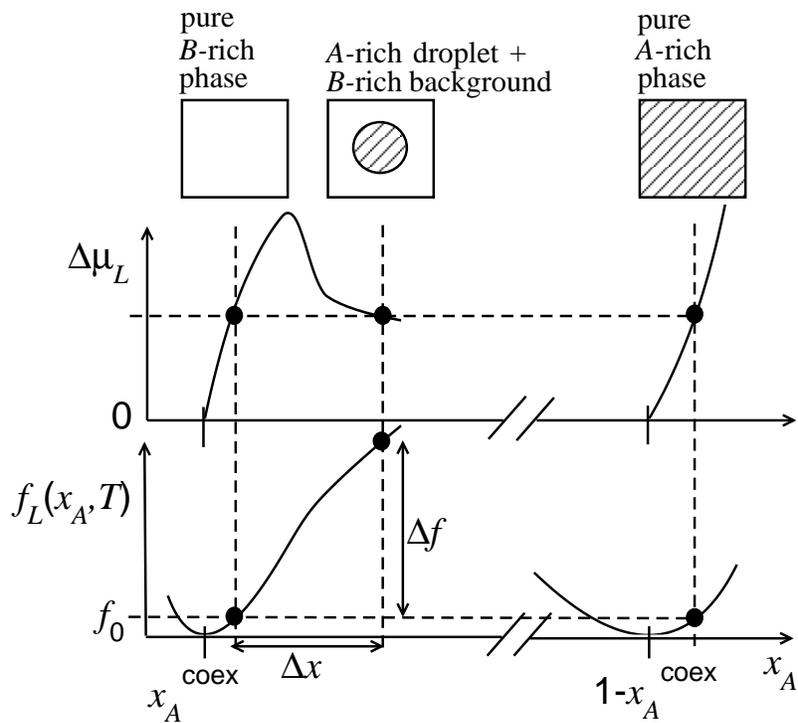}
\caption{Schematic explanation of how the estimation of the functions
$\Delta \mu_L(x_A,T)$ and $f_L (x_A, T)$ together allows the
estimation of the concentration difference $\Delta x$ and free
energy difference $\Delta f$ due to a droplet.}\label{fig7}
\end{figure}

Fig.~\ref{fig7} now recalls the method by which the surface free
energy $F_s(R)$ of the droplet and its radius $R$ are extracted
\cite{39,55,56}. An A-rich droplet coexists with a B-rich
background having the same chemical potential as another
state of pure B-rich phase. The chemical potential of the latter
must be equal to those of the A-rich droplet. Use of Eq.(\ref{eq1})
implies that one takes the bulk properties of the droplet, as a 
definition, equal to those of this B-rich phase, including its
bulk free energy density $f_0$ as indicated in the lower part of the 
figure. The difference $\Delta f$ in Fig.\ref{fig7} simply is due to the
surface free energy of the droplet
\begin{equation} \label{eq12}
V\Delta f= 4 \pi R^2 \gamma_{AB} (R).
\end{equation}
On the other hand, from Eq.~(\ref{eq1}), or its analog for a
binary mixture, we can readily obtain $R$ from the concentration
difference $\Delta x$ between the respective states as
\begin{equation} \label{eq13}
\Delta x= (1-2 x_A^{\rm coex}) (4 \pi R^3/3 L^3),
\end{equation}
where $x^{\rm coex}_A$ is the concentration of the pure B-rich
phase {on} the coexistence curve. Of course, for more precise estimation, $1-2 x^{\rm
coex}_A$ in Eq.(\ref{eq13}) should be replaced by the difference between the
compositions of the pure A-rich and B-rich phases at the
considered value of $\Delta \mu$, but in practice Eq.~(\ref{eq13})
is an excellent approximation. Estimating hence $\Delta f$ and
$\Delta x$ from the simulation, Eqs.~(\ref{eq12}),~(\ref{eq13})
yield the corresponding values $\gamma_{AB}(R)$ and $R$.
One clearly sees that both $\Delta {\mu_L}
(x_A,T)$ and $f_L(x_A,T)$ depend on the linear dimension $L$ of the
system, which also appears explicitly in Eq.~(\ref{eq13}).
However, the interpretation in terms of $\gamma_{AB}(R)$ makes
only sense if this dependence on $L$ completely drops out - at
least to a very good approximation. 

\begin{figure}[htb]
\includegraphics*[width=0.6\textwidth]{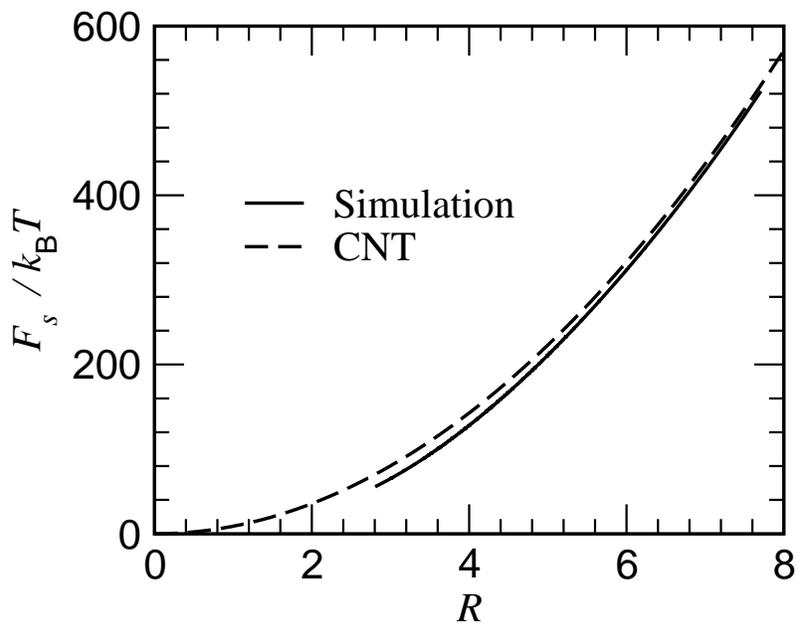}
\caption{Plot of $F_S/k_BT$ of spherical A-rich droplet at
$T=1$ for the binary symmetric LJ mixture of
Fig.~\ref{fig1}. The description in terms of the capillarity
approximation of classical nucleation theory (CNT) is shown as a
broken curve, using $\gamma_{AB}(\infty)=0.722$ as obtained in Fig.~\ref{fig3}.
The full curve is a superposition of independent
simulation results for $L=12$, 16,18,20,22 and 24,
where a running averaging was done using the combined data set.}\label{fig8}
\end{figure}

Fig.~\ref{fig8} presents the
resulting plot of $F_S(R)= 4 \pi R^2 \gamma_{AB} (R)$ vs. $R$ and
compares it with the capillarity approximation, $F_S(R)= 4 \pi R^2
\gamma_{AB} (\infty)$. One sees that the latter is 
accurate when the resulting surface free energies are rather
large, up to 600 $k_BT$! According to the classical nucleation
theory (CNT) \cite{1,2,3,4}, the resulting barrier $F^*$ to be
overcome in a homogeneous nucleation event would be
$F^*=F_S(R^*)/3$, $R^*$ being the radius of a critical nucleus.
This implies that for $4 \leq R^* \leq 8$ barriers
$F^*/k_BT$ of about 30 to 200 result. On the other hand, one can already see from
Fig.~\ref{fig8}, however, that for very small nuclei the relative
deviation from CNT increases.

We now turn to the question, {whether this deviation can} be accounted for
in terms of Tolman's hypothesis, Eg.~(\ref{eq4}). To test for the
latter, one notes that a plot of $\gamma(\infty)/\gamma(R)$ should
be equal to $1 + 2 \delta/R$, i.e. a straight line when plotted
against $1/R$. However, using the data from Fig.~\ref{fig8} no
straight line vs. $1/R$ can be observed, while the data are
compatible with a straight line when plotted against $1/R^2$,
as seen in Fig.~\ref{fig9}. This result agrees with the general
expectation, discussed already in the introduction, that for
systems that exhibit a precise symmetry between the coexisting
phases the Tolman length as defined in Eq.~(\ref{eq3}) is exactly
zero \cite{22,26}. Our results shown in Fig.~\ref{fig9}, as well
as a related study for the Ising model \cite{56}, implies that
instead of Eq.~(\ref{eq4}) the curvature-dependence of the surface
tension $\gamma_{AB} (R)$ for symmetric mixtures can be described
by
\begin{equation} \label{eq14}
  {\gamma_{AB}(R)= \gamma _{AB} (\infty)/[1+ 2(\ell_s/R)^2]},
\end{equation}
where the characteristic length $\ell_s$ at low temperatures is
close to the LJ diameter $\sigma$. It remains a challenge for the
future to clarify the temperature dependence of this length as $T
\rightarrow T_c$, however. We speculate that in this limit $\ell_s$
simply becomes proportional to the correlation length $\xi$ of order
parameter fluctuations. {This is supported by  explicit mean--field
results for the rigidities $k$ and $\bar{k}$ \cite{Blo93} to which $\ell_s$ is related
through $\ell_s^2 =-(k+\bar{k}/2)/\gamma_{AB}(\infty)$ (see Eq.~(\ref{eq:gams})) .}

In Fig.~\ref{fig9} deliberately only the range $1/R^2 \leq
0.12$ is shown. It should be noted that our method to construct
$\gamma_{AB}(R)$ is well-defined in the limit $R \rightarrow
\infty$, but as $R$ gets small, systematic errors arise: one reason
is that there is a nonzero, albeit very small, probability
that a state containing a droplet (or bubble), such as shown in
Fig.~\ref{fig5}a, undergoes a fluctuation to cylindrical shape
(Fig.~\ref{fig5}b) or to a homogeneous phase (Fig.~\ref{fig7},
upper part, left most cartoon). These rare fluctuations are
completely negligible for large $L$ used in Fig.~\ref{fig9}
but should become important if linear dimensions of the order of a
few $\sigma$ were used, and such linear dimensions would be needed
if we were to continue the study of the behavior towards larger
$1/R^2$. In addition, artefacts due to the periodic
boundary condition are to be expected, when $L$ is only of the
order of a few $\sigma$: fluctuations on the right side of the
droplet would interact with fluctuations on the left side of its
periodic image. Note that droplets with small $R$ require the use
of small $L$, in order to ensure their stability. In this respect,
our method that does not restrict statistical fluctuations
beyond the use of periodic boundary conditions on the scale $L$,
differs from the density functional theory (DFT) where one can
consider the equilibrium of an arbitrarily small droplet (at the
top $F^*$ of the nucleation barrier) in constrained equilibrium
with surrounding metastable phase extending infinitely far away
from the droplet. Due to the mean field character of DFT, this
constrained equilibrium does not decay while a corresponding
computer simulation clearly would result in an unstable decaying
situation.

\section{Droplets vs. Bubbles in the Single-Component Lennard-Jones Fluid}

In this section, we focus on the vapor-liquid
transition of simple fluids using the LJ potential,
following the work of Schrader et al. \cite{39}. But here, in addition to
spherical droplets, we also study spherical vapor
bubbles as well as droplets and bubbles of cylindrical shape. While
simulations of droplets surrounded by vapor are truly abundant
(and such simulations have been attempted since decades \cite{2}),
other geometries have found little attention, so far, despite
their physical significance.

As in Ref. \cite{39} we use a simple truncated LJ potential,
\begin{eqnarray} \label{eq15}
u(r_{ij}) &=& 4 \varepsilon[(\sigma/r_{ij})^{12} - (\sigma/r_{ij})^6
+ C], \quad r \leq r_c =2.2^{1/6} \sigma,\nonumber\\
u(r_{ij}) &=& 0, r\geq r_c, \quad C=127/16384.
\end{eqnarray}
Thus, the potential is cut at twice the distance of the
minimum, and the constant $C$ is chosen such that $u(r_{ij})$ is
continuous for $r_{ij}=r_c$. Two temperatures 
$T=0.68 T_c$ and $T=0.78 T_c$ ($T_c=0.999$) were studied with the linear dimension
of the simulation box varying from $L=11.3$ to $22.5$.
The accuracy of the estimation
of the excess free energy hump $f_L(\rho,T)$, $\rho$ ($=N/V$) now being 
the particle density, for large $L$ is
enhanced by applying the Wang-Landau method \cite{60} in addition
to successive umbrella sampling. 
Simulation method and the data analysis \cite{39} are rather similar to the
description provided in the previous section.

As an example, from the ``raw data'' obtained from these simulations,
Figs.~\ref{fig10} and \ref{fig11} show the free energy hump $\Delta
f(\rho, T)/k_BT$ at the chemical potential $\mu=\mu_{\rm coex}$
yielding phase coexistence between uniform saturated vapor and
liquid, as well as the derivative $\mu_L (\rho)/k_BT=(\partial
(f_L(\rho, T)/k_BT)/\partial \rho)_T$, for $T=0.78 T_c$. Similar
to the binary LJ mixture, the flat parts of $f_L(\rho)$ for
$\rho$ around 0.35, where $\Delta \mu_L(\rho)=0$, can be used to
find the vapor-liquid interfacial free energies $\gamma_{vl}(\infty)$ by an
extrapolation to $L \rightarrow \infty$
\cite{47,48,49,50,51,52,53}, in full analogy with Eq.~(\ref{eq10}).
Fig.~\ref{fig12} presents the counterpart of Fig.~\ref{fig3} in the present case, for $T=0.78T_c$.

The data shown in Figs.~\ref{fig10} and \ref{fig11}
are analyzed in an analogous manner as explained in the context of
Fig.~\ref{fig7}. 
The extension of this
method to the case of bubbles is completely straightforward: one
simply uses the part of the $\mu_L(\rho)$ and $f_L(\rho)$ curves
near $\rho=0.6$ rather than near $\rho=0.1$ in
Figs.~\ref{fig10},~\ref{fig11}. While traditionally droplets have
been identified as clusters of connected particles (e.g.
\cite{32}), such methods are not at all straightforward to
generalize in order to identify bubbles: the present thermodynamic method clearly has an
advantage here. Also the extension to cylindrical surfaces
is straightforward - we simply have to replace Eq.~(\ref{eq12}) by
an expression involving the surface area $2 \pi RL$ of a cylinder
surface,
\begin{equation} \label{eq17}
V\Delta f^{\rm cyl} = 2 \pi RL \gamma_{vl} ^{\rm cyl} (R),
\end{equation}
and Eq.~(\ref{eq13}) becomes modified by an expression containing the
volume of the cylinder, $\pi R^2 L$, rather than that of the
sphere, $4 \pi R^3/3$, so that
\begin{equation} \label{eq18}
\Delta x=(\rho _\ell - \rho_v)\pi R^2 /L^2.
\end{equation}

Fig.~\ref{fig13} presents the counterpart of Fig.~\ref{fig8}
for the single-component fluid, plotting $F_S/k_BT$ vs. the sphere radius, both
for droplets and bubbles, and compares them to the capillarity
approximation (CNT) where, of course, there is no difference between
droplets and bubbles. Indeed, we see that CNT
overestimates the correct surface free energies somewhat, and
$F_S(R)$ for bubbles falls clearly below the result for droplets
(such a difference {cannot occur in symmetric models}).
Fig.~\ref{fig14} presents the results for cylindrical geometries.

Figs.~\ref{fig15} and \ref{fig16} show our attempts to extract a
Tolman length from the
results of $\gamma_{vl}(R)$ for both droplets
and bubbles with spherical as well as cylindrical {shapes}. 
Motivated by the result for symmetric systems,
Eq.~(\ref{eq14}), where a quadratic term in $1/R$ is present while the
linear term being absent for symmetry reasons, we now assume the
presence of both linear and quadratic terms and
attempt to fit data to the forms:
{
\begin{eqnarray}\label{eq19}
\gamma_{vl} (\infty) /\gamma_{vl}(R) &=& 1 + 2 \delta /R + 2 (\ell_s/R)^2, \;
\textrm{spherical droplet}, \nonumber\\
\gamma_{vl} (\infty) /\gamma_{vl}(R) &=& 1 - 2 \delta /R + 2 (\ell_s/R)^2, \;
\textrm{spherical bubble}, \nonumber\\
\gamma_{vl} (\infty) /\gamma_{vl}(R) &=& 1 +  \delta /R + 2 (\ell_c/R)^2, \;
\textrm{cylindrical droplet}, \nonumber\\
\gamma_{vl} (\infty) /\gamma_{vl}(R) &=& 1 -  \delta /R + 2 (\ell_c/R)^2, \;
\textrm{cylindrical bubble}.
\end{eqnarray}
}
Eqs.~(\ref{eq19}) assume that in the leading order, droplets and bubbles, where
one goes from a convex to a concave interface,
just differ by a change of sign. 
{
The lengths $\ell_s$ and $\ell_c$ are related to
the rigidities $k$ and $\bar{k}$ introduced in Eqs.~(\ref{eq:gams}) and (\ref{eq:gamc})
by 
\bea
  \ell_s^2 &=& 2 \delta^2 - \frac{2k+\bar{k}}{2\gamma_{vl}(\infty)} \\
  \ell_c^2 &=& \frac{\delta^2}{2} - \frac{k}{4\gamma_{vl}(\infty)} \;.
\eea
}

Upon comparing
Eqs.~(\ref{eq19}) to the actual results obtained from unconstrained fits
to the simulation data, 
quoted in Figs.~\ref{fig15} and \ref{fig16}, we observe a clear indication of
linear correction, having a positive sign for bubble while being negative for droplet.
This tentatively implies a {negative} Tolman length of
order $\delta \approx -0.07 \pm 0.04$ for $T=0.68 T_c$ and $\delta
\approx -0.11 \pm 0.06$ for $T = 0.78 T_c$. Of course, the large
error bars which we simply extract from the scatter of the four
individual estimates (droplets and bubbles of spherical and cylindrical shapes) 
assuming that the symmetries postulated in
Eqs.~(\ref{eq19}) holds, are somewhat disappointing. Of course,
the possibility of systematic effects due to higher order terms
$\propto$ $R^{-3}, R^{-4}$ etc. cannot be ruled
out. However, the success of Eq.~(\ref{eq14}) in the symmetric case for a
similar range of $1/R$ (Fig.~\ref{fig9}) strengthens our expectation
that for the present case these higher order terms are still
negligible. The coefficient of the term $1/R^2$ is of order
unity, i.e., the {lengths $\ell_s$ and $\ell_c$ are} of the order of $\sigma$, as
found in the symmetric case. 
{From our fits we find negative bending rigidities $k$ with a magnitude of
about half a $k_B T$ whereas the rigidity $\bar{k}$ is consistent with zero.} 
We also note that the {lengths
$\delta$ and $\ell_s, \ell_c$} increase in absolute magnitude, with
increasing temperature. Thus, at least there is no qualitative
contradiction with the prediction that actually $\delta$ should
diverge as $T \rightarrow T_c$ \cite{37}. However, the smallness
of $\delta$ at $T=0.78 T_c$ precludes any hope that this possible
divergence might be probed by simulations.

We also note that ten Wolde and Frenkel \cite{32} in the analysis
of their simulation results for droplets (applying a rather
different method than in the present paper) suggested that
$\gamma_{vl}(\infty) / \gamma_{vl} (R) - 1 \propto 1/R^2$. This was motivated by
theoretical results of McGraw and Laaksonen \cite{63}, assuming
hence that $\delta = 0$ for a Lennard-Jones fluid. In view of our
result, that $\delta$ clearly is an order of magnitude smaller
than the length $\ell_s$ controlling the magnitude of the $R^{-2}$
term, it is understandable why ten Wolde and Frenkel missed the
existence of the Tolman correction. The strong point
of the present work is the proof of a clear difference between
$\gamma_{vl}(R)$ for droplets and bubbles, which is inconsistent with
\cite{63}. Figs.~\ref{fig13},\ref{fig14},\ref{fig15},\ref{fig16} give very clear evidence for the
existence of this difference, which according to Eq.~(\ref{eq19}) should
be
\begin{eqnarray}\label{eq20}
\Delta \gamma & \equiv & [\gamma_{vl} (\infty)/\gamma_{vl}
(R)]_{\textrm{bubbles}} - [\gamma_{vl} (\infty)/\gamma_{vl} (R)]
_{\textrm{droplets}}= 4 \delta /R + O(R^{-3}),\;
\textrm{spheres} \nonumber\\
\Delta \gamma & \equiv & [\gamma_{vl} (\infty)/\gamma_{vl}
(R)]_{\textrm{bubbles}} - [\gamma_{vl} (\infty)/\gamma_{vl} (R)]
_{\textrm{droplets}}= 2 \delta /R + O(R^{-3}),\;
\textrm{cylinders}
\end{eqnarray}
Our numerical data would yield, for $T=0.78 T_c$,
\begin{eqnarray}\label{eq21}
\Delta \gamma &=& 0.3811 (1 /R) + 0.266 (1 /R)^2, \quad
\textrm{spheres}, \nonumber\\
\Delta \gamma &=& 0.2386 (1 /R) - 0.172 (1 /R)^2, \quad
\textrm{cylinders}.
\end{eqnarray}
The small value
of the coefficient of $1/R^2$, compared to the individual
contributions coming from droplets and bubbles, is consistent with
the expected missing quadratic term in Eq. (\ref{eq20}). The
coefficient of the linear term for spheres, on the other hand, is a factor of 1.6
larger than that for cylinders and is again consistent with the expected theoretical factor 2,
{which is} gratifying.
Also, our estimates for $\delta$ certainly are compatible with the
result of van Giessen and Blokhuis \cite{38} $\delta = - 0.10 \pm
0.02$.

\section{RESULTS FROM DENSITY FUNCTIONAL THEORY}

For the one--component Lennard--Jones fluid, specified by the interaction
potential in Eq.~(\ref{eq15}), we have performed density functional calculations
for the metastable bubble and droplets in spherical geometry.
Here, we will treat the attractive part of the potential in a mean--field fashion
which is known to lack quantitative agreement with simulations for the phase diagram
or the liquid--vapor surface tension. However, the density functional approach
allows us to study also the limit of large droplet radii to check the validity
of the asymptotic expressions (\ref{eq3}) and (\ref{eq4}), thus complementing our simulation results.

As usual, the free energy functional of the fluid is split into an ideal part and an excess part,
\bea
  {\cal F}[\rho] &=& {\cal F}^{\rm id} [\rho] +  {\cal F}^{\rm ex}[\rho]
\eea
with the exact form of the ideal part given by
\bea
  {\cal F}^{\rm id}[\rho] &=& \int d^3r\,f^{\rm id}(\vect r) = \int d^3r\, \rho(\vect r )\left(
 \ln[ \rho(\vect r) \Lambda^3] -1 \right)\;.
\eea
Here, $\Lambda$ is the de--Broglie wavelength. The excess part is split into a reference hard--sphere
part and an attractive part:
\bea
  {\cal F^{\rm ex}}[\rho] &=& {\cal F^{\rm hs}}[\rho] + {\cal F^{\rm att}}[\rho] \;.
\eea
For the hard--sphere part we fix the reference hard--sphere diameter to $\sigma$, for simplicity, and
employ fundamental measure functionals which are known to be very precise in various circumstances
(for recent reviews see Refs.~\cite{Tar08,Rot10}).
\bea
 \label{eq:fhs}
 \beta {\cal F}^{\rm hs} &=& \int d\vect r \, \Phi( \{\vect n[\rho (\vect r)]\} ) \;,  \\
   \Phi( \{\vect n[\rho (\vect r)]\} ) &=&   -n_0\,\ln(1-n_3) +
      \frac{n_1 n_2-\vect n_1 \cdot \vect n_2}{1-n_3} +
      \varphi(n_3)\;\frac{n_2^3-3n_2\, \vect n_2\cdot \vect n_2}{24\pi(1-n_3)^2}\;.
   \nonumber
\eea
Here, $\Phi$ is  a
free energy density which is a function of a set of weighted densities
$\{\vect n (\vect r)\} = \{ n_0,n_1,n_2,n_3,\vect n_1,\vect n_2\}$
with four scalar and two vector densities. These are related to the
density profile $\rho(\vect r)$ by
$n_\alpha{(\vect r)} =  \int d\vect r' \rho(\vect r') \, w^\alpha(\vect r- \vect r')$.
The weight functions,
$\{\vect w(\vect r)\} = \{w^0,w^1,w^2,w^3, \vect w^1,\vect w^2\}$,
depend on the hard sphere radius $R=\sigma/2$
as follows:
\bea
 w^3 = \theta(R-|\vect r|)\;, \qquad
 w^2 = \delta(R-|\vect r|)\;, \qquad w^1 = \frac{w^2} {4\pi R}\;,
  \qquad w^0 = \frac{w^2}{4\pi R^2}\;,
  \nonumber\\
 \vect w^2 =\frac{\vect r}{|\vect r|}\delta(R-|\vect r|)\;,
 \qquad \vect w^1 = \frac{\vect w^2}{4\pi R} \;.
\eea
The reference hard--sphere free energy functional in Eq.~(\ref{eq:fhs}) is completed upon specification
of the function $\varphi(n_3)$. With the choice $\varphi = 1$
we obtain the original Rosenfeld (RF) functional \cite{Ros89}, consistent with the Percus--Yevick
equation of state. Upon setting
\bea
 \label{eq:fwb}
 \varphi & = & 1 - \frac{-2n_3+3n_3^2 - 2(1-n_3 )^2 \ln(1-n_3 ) }
                          {3 n_3^2}
\eea
we obtain the White Bear (WB) functional \cite{Rot02,Yu02}, consistent with the quasi--exact
Carnahan--Starling equation of state. For the attractive part of the free energy functional
we employ a mean--field (RPA) approximation motivated by the WCA potential separation \cite{WCA71}:
\bea
  {\cal F}^{\rm att}[\rho] &=& {\frac{1}{2}} \int d\vect r \int d\vect r' \rho(\vect r) \rho(\vect r')\;
    w(|\vect r - \vect r'|) \;, \\
  w(|\vect r|) &=& \left\{  \begin{matrix} u(r_{\rm min}) & \qquad (r \le r_{\rm min}) \\
                                           u(r) & \qquad (r>r_{\rm min})
                            \end{matrix}  \right. \;.
\eea
Here, $r_{\rm min}=2^{1/6}\sigma$ denotes the minimum location of the truncated Lennard--Jones
potential $u$ defined in Eq.~(\ref{eq15}).

The phase diagram of the model can be calculated easily by evaluating the free energy functional
for constant bulk densities and employing the Maxwell construction. The critical temperature,
$T_c^{\rm mf}/\epsilon \approx 1.15 $ is about 15\% too large, a defect well--known in mean--field
models. The discrepancies
in the phase diagram between our mean--field model and simulation can be significantly reduced
if the reference hard--sphere functional is used to generate a closure for the Ornstein--Zernike
equation \cite{Oet05,Aya09} and the reference hard--sphere diameter is appropriately determined
through a condition on minimal bulk free energy. However, this approach generates a lot more numerical work
and the results regarding the surface tension of bubbles and droplets remain qualitatively unchanged
(see below).

Planar and spherical surface tensions are obtained by extremizing the grand potential
\bea
  \Omega[\rho] &=& {\cal F}[\rho] -\mu \int \rho(\vect r) d\vect r
\eea
in the appropriate geometry,
which leads to the following equation for the  equilibrium density profile $\req(\vect r)$
\bea
 \label{eq:rhoeq_main}
  \ln \frac{\req(\vect r)}{\rho_0}  =
   -\beta\frac{\delta {\cal F}^{\rm ex}} {\delta \rho(\vect r)}[\req]+ \beta\mex  \;,
\eea
where the asymptotic bulk density is given by $\rho_0$ and $\mu\equiv\mu(\rho_0)$
is the corresponding chemical potential ($\mex$ denotes its excess part).
In planar geometry, the grand potential extremum is indeed a minimum if the boundary
conditions for $\req^{\rm planar}(z)$  are such that the coexisting
vapor density $\rho_{\rm v}$ is enforced as $z\to -\infty$ and the corresponding coexisting
liquid density $\rho_{\rm l}$ is enforced as $z\to \infty$. The planar surface tension thus becomes
\bea
  \gamma_{vl} &=& \Omega'[\req^{\rm planar}(z)] - \Omega'[\rho_{\rm v}] = \Omega'[\req^{\rm planar}(z)]  + p(\rho_{\rm v}) L \;,
\eea
where $\Omega'=\Omega/A$ is the aerial grand potential density, and $L$ is the length of our numerical box in $z$--direction.
In spherical geometry, the asymptotic bulk density $\rho_0=\lim_{r\to\infty}\req^{\rm sph}$ has to be chosen either as an
oversaturated liquid (for a vapor bubble) or an oversaturated vapor (for a liquid bubble). The
corresponding extremal point of the grand potential is a saddle point which makes the  extremization
a bit more cumbersome. Simple iteration schemes will always diverge (``evaporating'' the droplet)
after some initial period of convergence.
Therefore we have chosen to adopt a Picard iteration scheme along with suitable radial shifts of the whole density profile
each time the Picard iteration starts to diverge. This procedure is repeated until the necessary radial shifts
are below the resolution of our radial grid which is 0.001 $\sigma$. From the equilibrium profile  $\req^{\rm sph}(r)$
the equimolar droplet radius $R$ is determined by the condition of no net adsorption. For the associated
surface tension it is necessary to introduce the asymptotic ``inner'' density of the drop $\rho_1 \not= \rho_0$ which
is given by the condition $\mu(\rho_1)=\mu(\rho_0)$. (For small drops, the actual density in the center is not necessarily
equal to $\rho_1$.) The excess pressure within the drop is defined via our mean--field equation of state as
$\Delta p = p(\rho_1)-p(\rho_0)$. Furthermore it is convenient to introduce the excess grand potential of the drop
by
\bea
  \Delta \Omega = \Omega[\req^{\rm sph}(r)] + \frac{4\pi}{3}L_{\rm sph}^3\, p_0
\eea
where $L_{\rm sph}$ is the radius of our spherical numerical box.
This excess grand potential constitutes the nucleation barrier at pressure $p_0$.
The (equimolar) surface tension then becomes
\bea
 \label{eq:gamma_e_def}
   \gamma_{vl}(R) = \frac{1}{4\pi R^2} \left( \Delta\Omega + \frac{4\pi}{3} R^3 \,\Delta p \right) \;,
\eea
The radius $R_p$ of the ``surface of tension''
is defined by the Laplace condition $\Delta p = 2\gamma_{vl}(R_p)/R_p$. Little algebra leads to the relations:
\bea
   R_p &=&  \left( \frac{3\Delta\Omega}{2\pi\Delta p} \right)^{\frac{1}{3}} \;, \\
 \gamma_{vl}(R_p) &=& \left( \frac{3\Delta\Omega\Delta p^2}{16\pi} \right)^{\frac{1}{3}} \;.
\eea

\begin{table}
 \begin{tabular}{lllllll}
   \hline \hline \\
   $T/T_c$ & $\rho_v\sigma^3$ & $\rho_l\sigma^3$ & $\gamma_{vl}\sigma^2/\epsilon$ & $\rho_v\sigma^3$ & $\rho_l\sigma^3$ & $\gamma_{vl}\sigma^2/\epsilon$ \\
               & \multicolumn{3}{c}{(DFT)} & \multicolumn{3}{c}{(Simulation)} \\ \hline
   0.68$\quad$ & 0.005$\quad$ & 0.79$\quad$ & 0.56$\qquad$ & 0.010$\quad$ & 0.77$\quad$ & 0.47 \\
   0.78 & 0.010 & 0.71 & 0.39 & 0.027 & 0.71  & 0.29 \\ \hline \hline
 \end{tabular}
 \caption{\label{tab:coex} Comparison between DFT and simulation results for the coexistence densities and liquid--vapor surface tension
 at the two investigated temperatures. }
\end{table}

We report numerical results using the WB functional as the reference hard--sphere functional
(see Eqs.~(\ref{eq:fhs}) and (\ref{eq:fwb})). Coexistence densities and planar surface tensions for the
two investigated temperatures $T/T_c=0.68$ and 0.78 are given in Tab.~\ref{tab:coex}. Since we are quite
far from the critical point, the match in the coexistence densities between DFT and simulation is quite
reasonable. There is, however, a mismatch in the planar surface tension of about 25\%.
Next we evaluated the surface tensions of bubbles and droplets for radii between 2 and approximately
200 $\sigma$. This allows us to test the asymptotic expressions for the Tolman length given
in Eqs.~(\ref{eq3}) and (\ref{eq4}). The results for $\delta(R)=R-R_p(R)$ and $\gamma_{vl}/\gamma_{vl}(R)-1$ are
shown in Figs.~\ref{fig:DFT_T0.78} (for $T/T_c=0.78$) and \ref{fig:DFT_T0.68} (for $T/T_c=0.68$).
One can see that through a linear fit to $\delta$ as a function of $1/R$ one obtains the Tolman length to
a good precision: $\delta = -0.127 \pm 0.002$ for $T/T_c=0.78$ and $\delta = -0.122  \pm 0.002$ for $T/T_c=0.68$.
The linear terms in the quadratic fits to $\gamma_{vl}/\gamma_{vl}(R)-1$ (whose modulus should equal $2\delta$) are in agreement with
these results although here
the discrepancy between the bubble and droplet results are a bit larger:
$\delta = -0.129 \pm 0.002$ for $T/T_c=0.78$ and
 $\delta = -0.125  \pm 0.008$ for $T/T_c=0.68$.
Overall these results for the Tolman length are consistent with the simulation results of the previous section, as
is the qualitative behavior of $\gamma_{vl}/\gamma_{vl}(R)-1$. The most notable discrepancy arises in the fit coefficient
of the quadratic term which is about a factor 4 smaller in DFT, and thus the variation of $\gamma_{vl}(R)$ with $R$ is much
less pronounced  in DFT than it is in the simulation results. To understand this discrepancy partly,
note that in the determination of $\gamma_{vl}(R)$ in DFT (Eq.~(\ref{eq:gamma_e_def}) we use the
mean--field bulk equation of state in the metastable domain in order to subtract the bulk grand potential
of the droplet. This can be expected to lead to different results compared to the simulations where
the free energy density $f_L(\rho,T)$ of a finite system (with box length $L$ and density $\rho$, exhibiting
a stable droplet) is determined and used to subtract the bulk grand potential
of the droplet.
However, one can also expect that the correlations inside the droplet are not adequately captured by the
mean--field treatment.

Finally we compare our DFT approach to existing DFT approaches in the literature. All DFT results have been obtained
using a mean--field {\em ansatz} for the effect of attractions and differ mainly by the treatment of the
hard repulsions \cite{22,28,29,30,Kog98,Nap01,34,38}. Differences in the latter manifest themselves in the absolute numbers for the
surface tension whereas the results for the reduced quantity $\gamma_{vl}/\gamma_{vl}(R)$ are basically unaffected
(see e.g. Refs.~\cite{Kog98,Nap01}). (Differences arise for smaller $R$ upon using different subtraction
schemes for the bulk grand potential of the droplet). All DFT results for Lennard--Jones type liquids
(with differing cut--offs, applied
also at various temperatures)
have so far predicted a negative Tolman length. Here we have confirmed this finding by considering also significantly
larger droplets and by checking the consistency of the Tolman length extraction from both vapor bubbles and
liquid droplets.

\section{Conclusions}

In the present work a computer simulation approach to explore the
surface free energy of curved interfaces is presented and applied
to perform a comparative study of droplets of the minority phase
in an unmixed symmetrical binary Lennard-Jones mixture and of both
droplets and bubbles in the  two-phase coexistence region of a
simple Lennard Jones fluid. In the latter case, both droplets and
bubbles of spherical and
cylindrical shapes have
been analyzed with the basic finding of a 
systematic difference between the
curvature-dependent surface free energy of droplets and bubbles.
Finally, also a density functional treatment of
both droplets and bubbles in simple fluids is presented that
supports the latter observation.

\begin{figure}[htb]
\includegraphics*[width=0.6\textwidth]{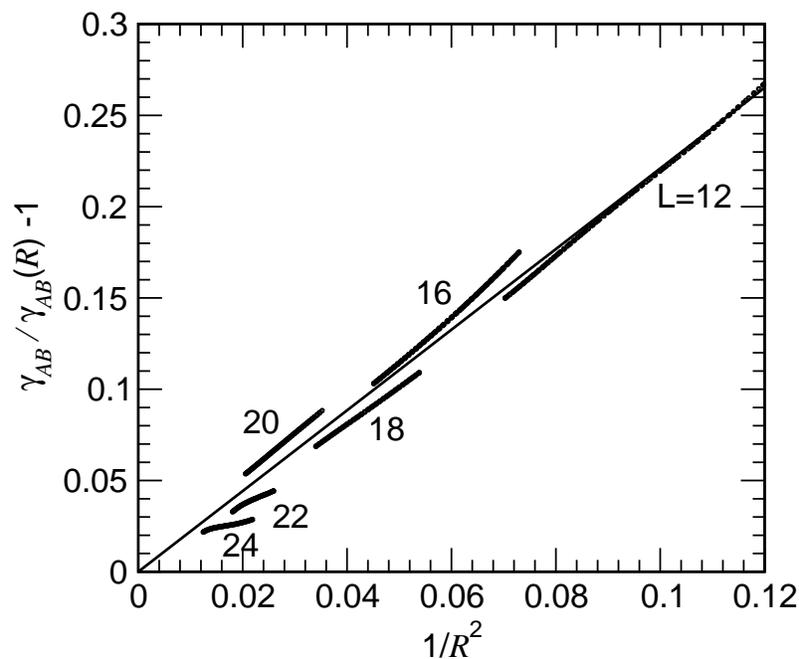}
\caption{Plot of $\gamma_{AB}(\infty)/\gamma_{AB} (R)-1$ versus
$1/R^2$. Here $\gamma_{AB}(\infty)$ is taken from Fig.~\ref{fig3},
while $\gamma_{AB}(R)\equiv F_S(R)/4 \pi R^2$ is estimated using
Eq.~(\ref{eq12}). Ideally the estimates obtained from different
values of $L$ should superimpose on a single curve. The scatter
between the curves for different values of $L$ is due to residual
statistical errors. The thin straight line is a fit function giving
$\gamma_{AB}(\infty)/\gamma_{AB}(R)-1\simeq 2.2/R^2.$}\label{fig9}
\end{figure}

\begin{figure}[htb] 
\includegraphics*[width=0.6\textwidth]{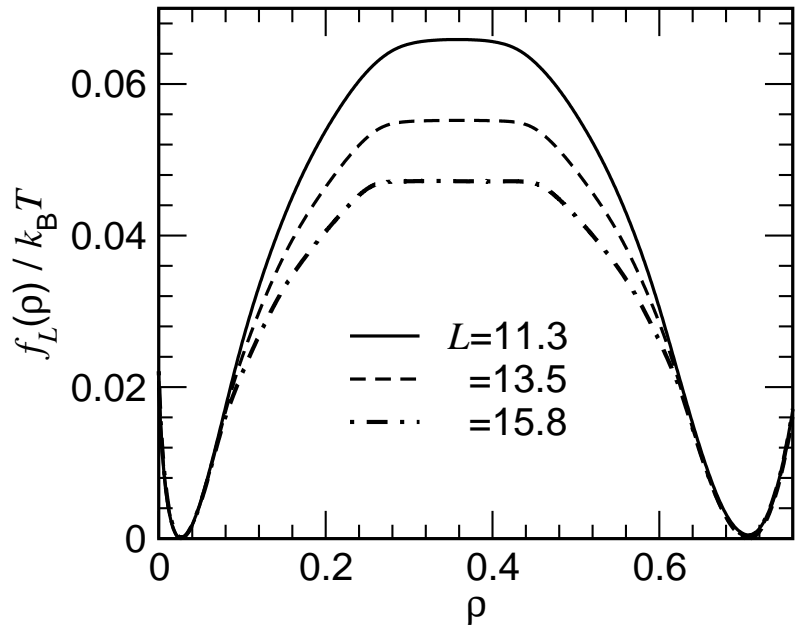}
\caption{Effective free energy density $f_L(\rho)/k_BT$ of the single-component
Lennard-Jones fluid at $T=0.78T_c$ plotted vs. density $\rho$ for
3 values of $L$, as indicated in the figure.}
\label{fig10}
\end{figure}

\begin{figure}[htb] 
\includegraphics*[width=0.6\textwidth]{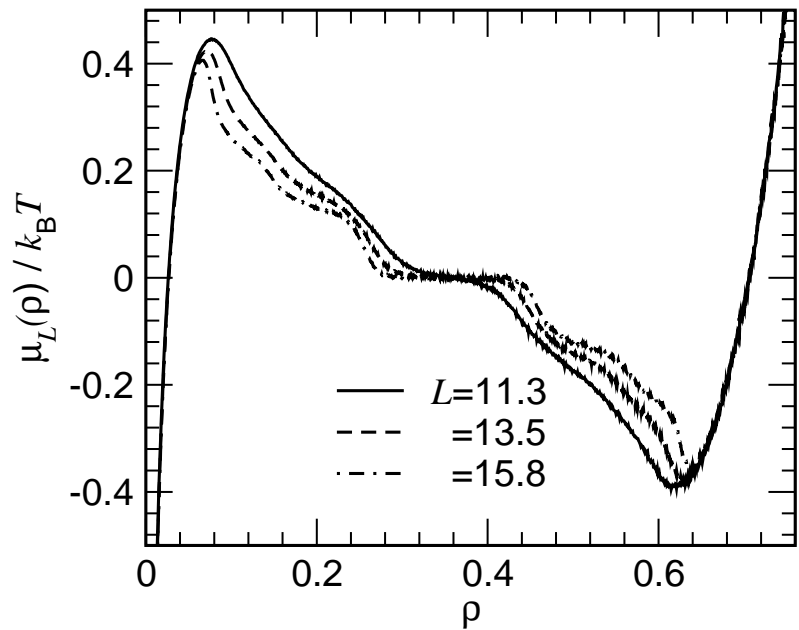}
\caption{Plot of $\mu_L(\rho)/k_BT$ for the one-component
Lennard-Jones fluid as a function of $\rho$, at $T=0.78T_c$, for 3
values of $L$, as indicated.}\label{fig11}
\end{figure}

\begin{figure}[htb] 
\includegraphics*[width=0.6\textwidth]{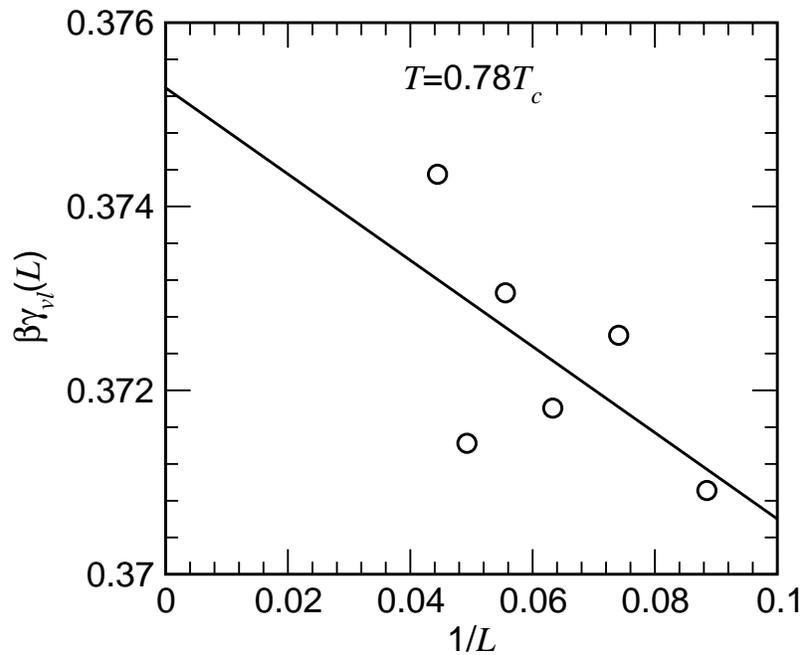}
\caption{Extrapolation of $\beta\gamma_{vl}(L)$ as a
function of $1/L$ for the simple LJ fluid at $T=0.78T_c$, giving
$\beta\gamma_{vl}=\beta\gamma_{vl}(\infty)\simeq 0.375$. Similar exercise at
$T=0.68T_c$ gives $\beta\gamma_{vl}\simeq 0.685$.}
\label{fig12}
\end{figure}

For the symmetric binary Lennard-Jones mixture, such a
difference cannot occur, of course, since an interchange of A and
B in the two-phase coexistence turns the minority phase into the
majority phase, and a droplet becomes a bubble, or vice versa. As
a consequence, Eq.~(\ref{eq4}), with a
{nonzero} Tolmann length $\delta$, cannot apply, {furthermore} the dependence of
the surface tension $\gamma_{AB} (R)$ on the droplet radius of
curvature cannot contain any odd term in $1/R$,
since $R$ changes sign when a droplet is turned into a
bubble. We find that for temperatures far below criticality, the
simple formula $\gamma_{AB}(R)=\gamma_{AB} (\infty)/[1+2
(\ell_s/R)^2 ] $ \{Eq.~(\ref{eq14})\} is a very good representation of
our data, for radii $R$ down to about 3 Lennard-Jones diameters,
and the characteristic length $\ell_s$ is a bit longer than a
Lennard-Jones diameter (Fig.~\ref{fig9}).

\begin{figure}[htb] 
\includegraphics*[width=0.6\textwidth]{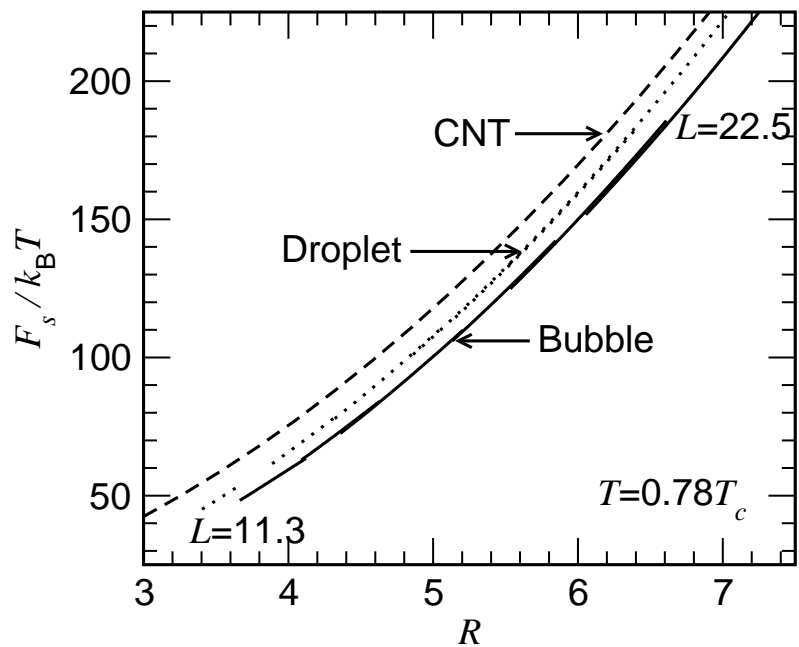}
\caption{Plots of $F_S/k_BT$ of spherical droplets and bubbles
for the one-component LJ fluid, at $T=0.78T_c$, as a function of sphere radius
$R$. The capillarity approximation (CNT), $F_S/k_BT=4 \pi R^2
\gamma_{vl}(\infty)$ is included, using the estimate of $\gamma (\infty)$ from
Fig.~\ref{fig12}.}\label{fig13}
\end{figure}

\begin{figure}[htb] 
\includegraphics*[width=0.6\textwidth]{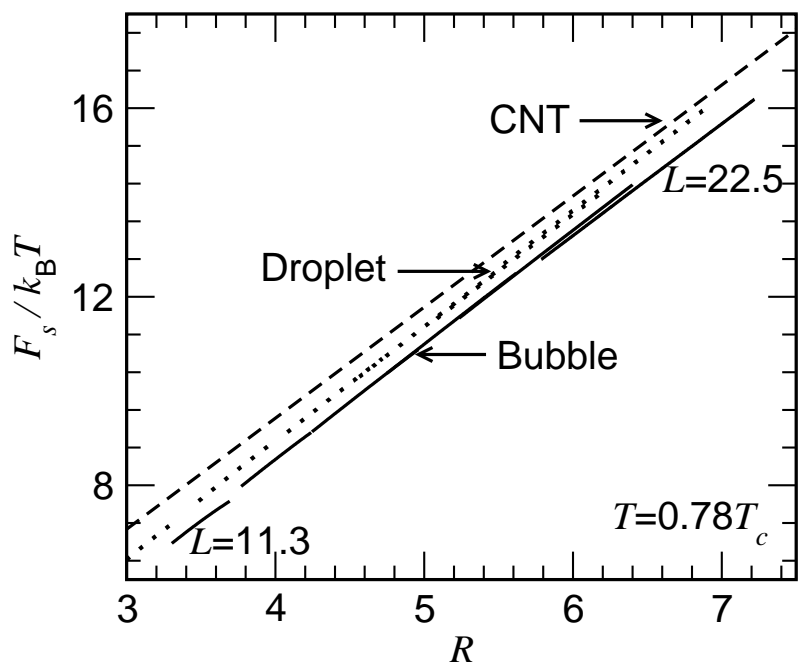}
\caption{Same as Fig.~\ref{fig13}, but for cylindrical droplets
and bubbles. Note that, here the $y$-axis corresponds to the
surface free energy per unit height of the cylinder.}\label{fig14}
\end{figure}

\begin{figure}[htb]
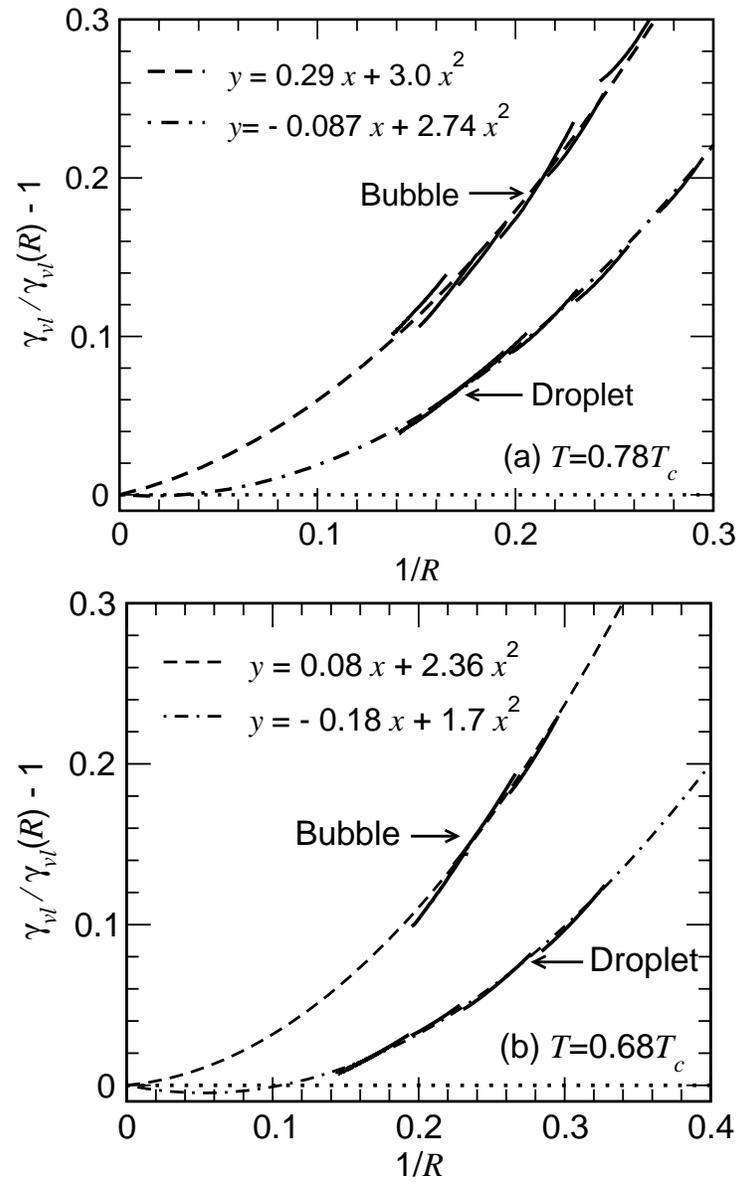
 
\includegraphics*[width=0.55\textwidth]{fig15a.eps}\\
\includegraphics*[width=0.55\textwidth]{fig15b.eps}
\caption{Plots of $\gamma_{vl} (R)/\gamma_{vl}(\infty)-1$ vs. $1/R$ for spherical
droplets and bubbles for the LJ fluid at (a) $T=0.78 T_c$ and (b)
$T=0.68 T_c$. Fits to the functional forms (\ref{eq19}) are included.
}\label{fig15}
\end{figure}

\begin{figure}[htb]
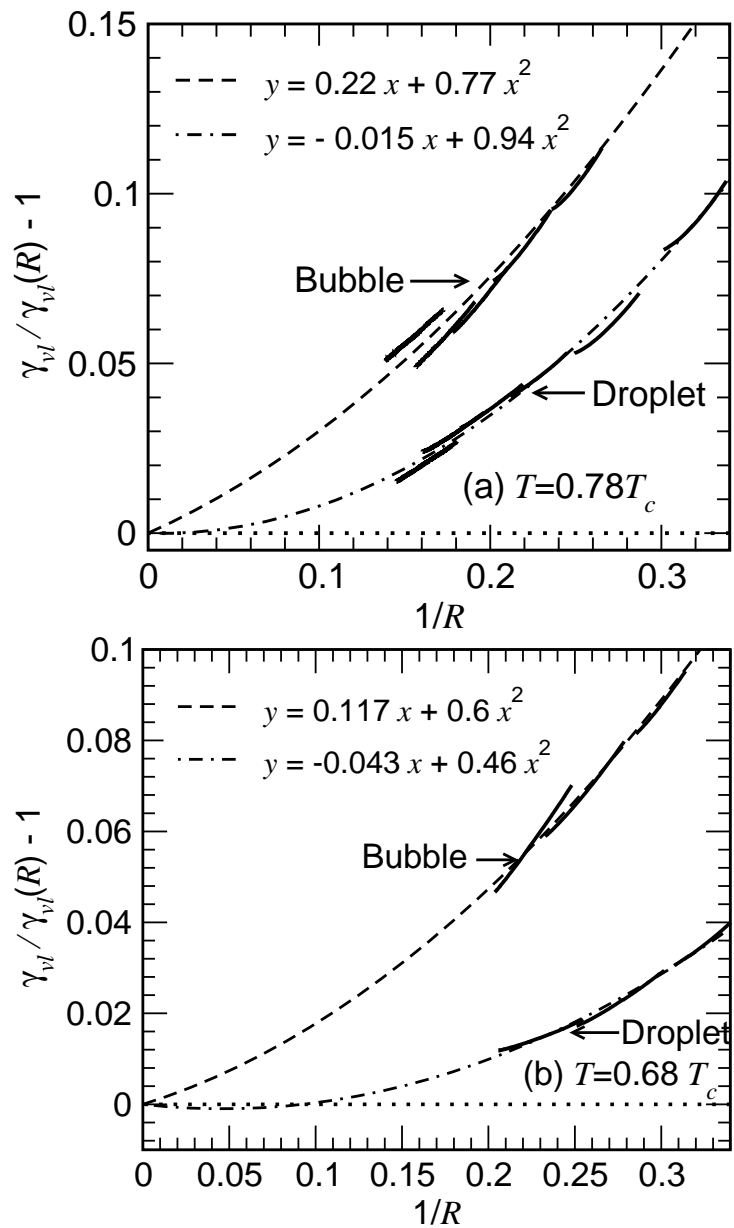
 
\includegraphics*[width=0.55\textwidth]{fig16a.eps}\\
\includegraphics*[width=0.55\textwidth]{fig16b.eps}
\caption{
Same as Fig.~\ref{fig15}, but for cylindrical droplets and
bubbles.}\label{fig16}
\end{figure}

\begin{figure}[htb]
\includegraphics*[width=0.55\textwidth]{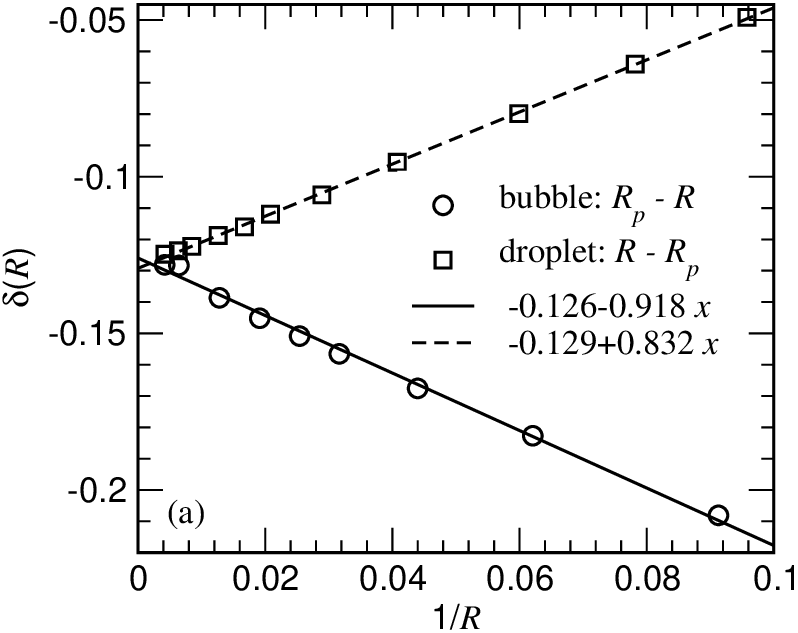}
\includegraphics*[width=0.55\textwidth]{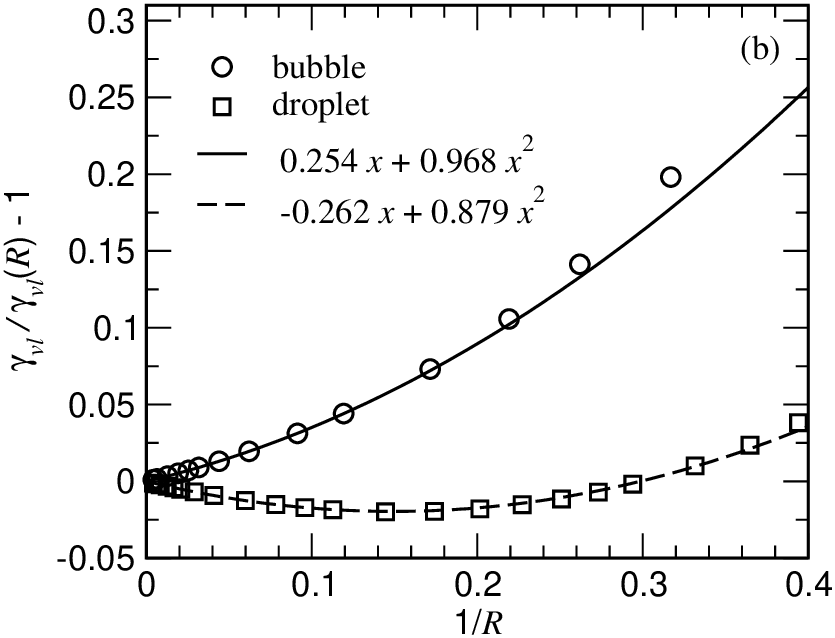}
\caption{ \label{fig:DFT_T0.78} (a) Radius--dependent Tolman length $\delta(R) = R-R_p$ for bubbles and droplets
with a corresponding linear fit in the range $1/R \in (0,0.1)$. (b) Surface tension ratio $\gamma_{vl}/\gamma_{vl}(R) -1$
as function of the equimolar radius $R$ for bubbles and droplets with a corresponding quadratic fit in the range
$1/R \in (0,0.1)$. All results are for the temperature $T=0.78T_c$.}
\end{figure}

\begin{figure}[htb]
\includegraphics*[width=0.55\textwidth]{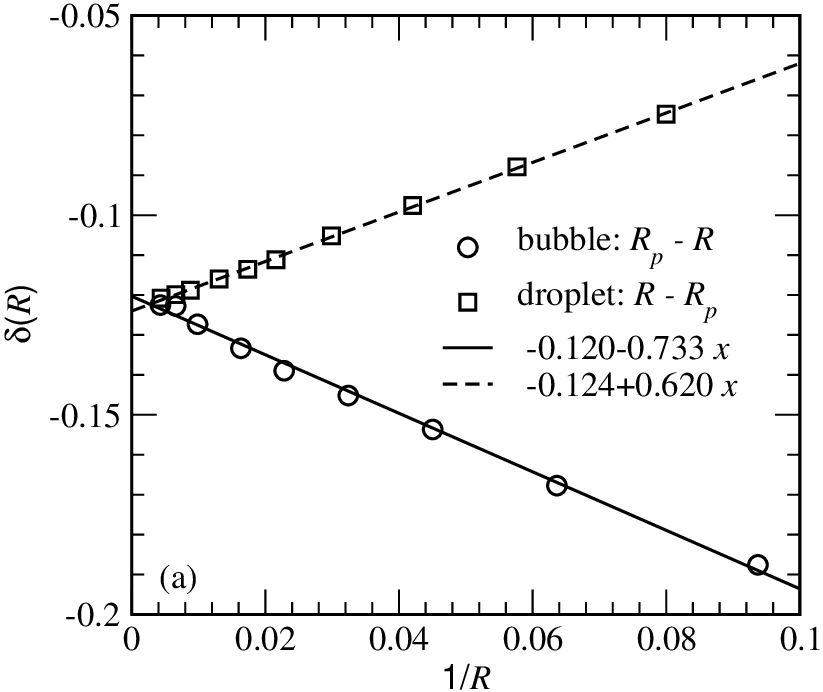}
\includegraphics*[width=0.55\textwidth]{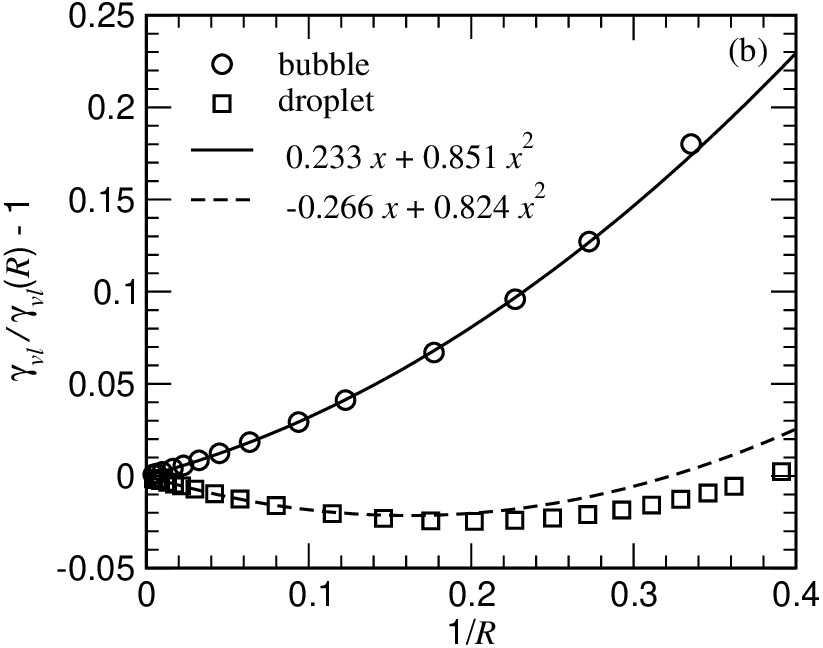}
\caption{ \label{fig:DFT_T0.68} (a) Radius--dependent Tolman length $\delta(R) = R-R_p$ for bubbles and droplets
with a corresponding linear fit in the range $1/R \in (0,0.1)$. (b) Surface tension ratio $\gamma_{vl}/\gamma_{vl}(R) -1$
as function of the equimolar radius $R$ for bubbles and droplets with a corresponding quadratic fit in the range
$1/R \in (0,0.1)$.  All results are for the temperature $T=0.68T_c$.  }
\end{figure}

The main result for the droplets and bubbles in the
simple Lennard-Jones fluid (which lacks the trivial
particle-hole symmetry of the
simplistic lattice gas model)
is the finding that
there is a significant difference between $\gamma (R)$ of droplets
and bubbles (Fig.~\ref{fig13},\ref{fig14},\ref{fig15},\ref{fig16}).
This shows that the conventional ``capillarity
approximation'' of the standard classical nucleation theory is not
strictly valid, since the approximation does not allow for any
such difference. Also theories such as those of McGraw and
Laaksonen \cite{63}, which imply that the Tolmann length $\delta$
is strictly zero and the difference $\gamma (\infty)-\gamma(R)$, to
leading order, is $1/R^2$, are hence inconsistent with our results.
We do find that this difference should be described both
by a linear term $(\delta/R)$ and a quadratic term $(\ell_s/R)^2$
[Eq.~(\ref{eq18})], while the length $\ell_s$ again is of the
order of the Lennard-Jones diameter $\sigma$, as for the symmetric
binary Lennard-Jones mixture, the Tolman length $\delta$ being an
order of magnitude smaller, $\delta \approx -0.16$. This order of
magnitude is compatible with the conclusion of some of the previous
work \cite{38,40}. Qualitatively, these conclusions are fully
confirmed by the density functional treatment (Sec. IV): note that
because of the mean-field character of the latter, it does not
precisely reproduce the equation of state of the model studied in
the simulation and so it would be premature to ask for a
quantitative ``fit'' of the simulation results
(Fig.~\ref{fig15},~\ref{fig16}) by the theory
(Figs.~\ref{fig:DFT_T0.78},~\ref{fig:DFT_T0.68}). However, the qualitative similarity is
striking, and the order-of-magnitude agreement between the
corresponding numbers is, of course, gratifying.

An important consequence of our results also is that deviations
from classical nucleation theory for bubble nucleation of vapor
should be distinctly larger than for droplet nucleation of liquid,
at comparable radius $R$. Classical nucleation theory
overestimates the nucleation barrier in the range of practical
interest.

Of course, it would be interesting to study the temperature
dependence of the lengths $\delta$ and $\ell_s,\ell_c$, and to extend our
study to other models of fluids. {This will be left to future
work.}

\underline{Acknowledgements:} This work was supported in part by
the Deutsche Forschungsgemeinschaft (DFG) 
{through the Collaborative Research Centre SFB-TR6 under grants No.
TR6/A5 and TR6/N01 and through the Priority Program SPP 1296
under grants No. Bi 314/19 and Schi 853/2.}
S.K.D. is grateful to the Institut
f\"ur Physik (Mainz) for the hospitality during his extended visits.
We are also grateful to the Zentrum f\"ur Datenverarbeitung (ZDV)
Mainz and the J\"ulich Supercomoputer Centre (JSC) for computer
time. One of us (K.B.) is grateful to C. Dellago, D. Frenkel, G.
Jackson and E.A. M\"uller for stimulating discussions.

\clearpage

\end{document}